\documentclass[useAMS,usenatbib]{mn2e}

\usepackage{graphics,epsfig}
\usepackage{graphicx}
\usepackage{amssymb}

\newcommand{\tspin}{T$_{\rm s}$ }
\newcommand{\tzeroone}{T$_{\rm 01}$ }
\newcommand{\HI}{H~{\sc i} }
\newcommand{\htwo}{H$_2$ }
\newcommand{\kms}{km~s$^{-1}$ }
\newcommand{\dV}{\mathrm{d} v}

\title[Temperature of the CNM]{A multiwavelength investigation of the 
                                              temperature of the cold neutral medium}
\author[N. Roy, J. N. Chengalur and R. Srianand]{Nirupam Roy$^{1}$
       \thanks{E-mail: nirupam@ncra.tifr.res.in~(NR);~chengalu@ncra.tifr.res.in~(JNC);~anand@iucaa.ernet.in~(RS)}, 
       Jayaram N. Chengalur$^{1}$\footnotemark[1] and Raghunathan Srianand$^{2}$\footnotemark[1]\\ 
       $^{1}$NCRA-TIFR, Post Bag 3, Ganeshkhind, Pune 411 007, India\\
       $^{2}$IUCAA, Post Bag 4, Ganeshkhind, Pune 411007, India}

\begin{document}
\date{Accepted 2005 October 10. Received 2005 October 5; in original form 2005 May 19}

\pagerange{\pageref{firstpage}--\pageref{lastpage}} \pubyear{2005}

\maketitle

\label{firstpage}

\begin{abstract}

       We present measurements of the \HI spin temperatures (T$_{\rm s}$) of the Cold Neutral Medium (CNM) 
towards radio sources that are closely aligned with stars for which published \htwo ortho-para temperatures 
(T$_{\rm 01}$) are available from UV observations. Our sample consists of 18 radio sources close to
16 nearby stars.  The transverse separation of the lines of sight of corresponding the UV and radio 
observations varies from 0.1 to 12.0 pc at the distance of the star. The ultraviolet (UV) measurements do not have
velocity information, so we use the velocities of low ionization species (e.g Na~{\sc i}/$\,$K~{\sc i}/$\,$C~{\sc i}) 
observed towards these same stars to make a plausible identification of the CNM corresponding to the \htwo 
absorption. We then find that \tzeroone and \tspin match 
within observational uncertainties for lines-of-sight with \htwo column density above 10$^{15.8}$ cm$^{-2}$, 
but deviate from each other below this threshold. This is  consistent with the expectation 
that in the CNM  \tspin tracks the kinetic temperature due to collisions  and that \tzeroone is 
driven towards the kinetic temperature by proton exchange reactions.

\end{abstract}


\section{Introduction}

        Physical conditions in the interstellar medium (ISM) have 
traditionally been studied using spectral lines from a variety of tracers 
including the 21~cm line and recombination lines of hydrogen in the radio 
regime, Lyman lines of H~{\sc i}, the Lyman and Werner bands of H$_2$ 
and the atomic fine-structure lines such as C~{\sc i} in the UV 
as well as a host of rotational lines from molecules in the mm wavelength 
regime. For example the kinetic temperature of the gas can be
determined using either the H~{\sc i} 21~cm line or the H$_2$ UV lines,
the pressure and cooling rates can be determined from the fine 
structure lines of C~{\sc i} and C~{\sc ii$^*$} etc. Since many of 
these tracers co-exist in the diffuse ISM, multi-wavelength 
observations would allow one to cross check 
different observational  techniques as well as to derive a more 
complete understanding of the physical state of the ISM.

        The gas temperature is a particularly well suited example of
a parameter that can, in principle, be determined by observations at a variety of
wavelengths. In the radio regime, the classical method consists  
of observing the H~{\sc i} 21-cm line in absorption towards a bright radio 
continuum source; this, in conjunction with observations of the 
emission spectrum along a nearby line of sight allows one to measure 
the spin temperature (T$_{\rm s}$) of the \HI \citep[see e.g.][for details]
{kulkarni}. While the \HI spin temperature, strictly speaking, 
characterizes the population distribution  between the two hyperfine 
levels of the hydrogen atom, it is often used as a proxy for the kinetic 
temperature of the gas. This is because, in high density regions, \tspin 
is expected to be tightly coupled to the kinetic 
temperature via collisions, while in low density regions, 
resonant scattering of Lyman-$\alpha$ photons again may couple the spin 
temperature to the kinetic temperature \citep {d}. UV observations of the 
Lyman and Werner bands of H$_2$ also allow one to determine the gas temperature. 
This is the so called ``ortho-para'' temperature (T$_{\rm 01}$).
The ortho-para temperature characterizes the population distribution between
the ortho and para forms of the H$_2$ molecule, and like the spin temperature,
it too is expected to be coupled to the kinetic temperature of the gas. This 
happens mainly via proton exchange collisions \citep*[e.g.][]{c}. In regions where
\HI and H$_2$ co-exist, one might hence expect that the temperatures derived
from radio (i.e. T$_{\rm s}$) and UV observations (i.e. T$_{\rm 01}$) should match. 
However, there have been very limited multi-wavelength studies of this 
sort. The main reason for this is probably that a given line of sight 
is rarely suitable for observations at more than one wavelength. For example, 
UV observations are generally made toward bright nearby stars which have 
no detectable radio emission. This rules out complimentary 21-cm studies along these 
lines of sight. Although a direct comparison of the two temperatures along a given 
line of sight is difficult it is still possible to compare the average properties 
of the two temperatures measured from different surveys. For example, using a large 
sample of UV spectra towards nearby stars, \citet*{a} found the mean value 
of \tzeroone to be $77 \pm 17$~K, which is in good agreement with the mean 
value of the 21-cm \tspin in CNM. This lends support to the argument that 
both these temperatures trace the kinetic temperature of the gas.

       The recent large scale high sensitivity and high angular
resolution radio surveys like the NRAO VLA Sky Survey \citep*[NVSS,][]{r} however do allow a way around the
problem of finding lines of sight suitable for both UV and radio 
observations by identifying radio sources that happen to be close to 
the line of sight to UV bright stars. These radio sources are generally 
too faint for the classical single dish emission absorption studies,
but are well suited for interferometers where the subtraction of the smooth 
background emission is automatically achieved. In this work we present the 
results of the Giant Metrewave Radio Telescope \citep*[GMRT,][]{s} 21-cm H~{\sc i} observations toward 18 radio sources. 

\section{Observation and data analysis}
\label{obs}

Our sample (drawn from the NVSS catalog), consists of 18 radio sources 
brighter than 100 mJy. For each source, we require that there should
be a star within $30'$, along the line of sight to which \tzeroone is 
available in \citet{a}. Our 18 radio sources correspond to 16 distinct 
stars, two stars having two corresponding radio sources. 
The details are summarized in Table~1. The columns in Table~1 are: (1)~the 
name of the optical source and (2)~the radio source, (3)~the distances to 
the star, (4)~the stellar co-ordinates, (5)~the co-ordinates of the 
corresponding radio source, (6)~the total \htwo column density from \citet{a}, 
(7)~the line of sight extinction, $E(B-V)$ from \citet{a}, (8)~the ortho-para 
temperature and its error bars \citep[from][as well as from our 
recalculations from their data, see below]{a}, (9)~the \HI 21-cm 
spin temperature (see section~\ref{temp} for details) and (10)~the 
fractional difference between \tzeroone and T$_{\rm s}$. 
It is clear from Table~1 that lines of sight in our sample 
span wide range of $E(B-V)$ and  N(H$_2$). The transverse separations 
of the lines  of sight of corresponding UV and radio observations at 
the distance of the stars ranges from 0.1 to 12.0 pc. 

For the 6 sight lines with N(H$_2$)$\le10^{17}$ cm$^{-2}$, no error 
bars for the column densities were available in \citet{a}. For these 
cases, we obtained the column densities of H$_2$ in J=0, and J=1 
rotational levels of ground vibrational level using Voigt profile 
fitted to the {\it Copernicus} archival data. Our estimated column 
densities agree well with that reported by \citet{a} and we use our 
computed errors as indicative values while calculating the excitation 
temperatures for these systems. 

The GMRT radio observations were conducted between July 07--10 2002.  
For all sources the observing frequency was 1419.4 MHz and the total 
bandwidth was 2~MHz with 128 spectral channels (i.e. a velocity 
resolution of $\sim 3.3$ km~s$^{-1}$). The total on-source time was 3--4 hour 
on each source. Scans on standard calibrators 
were used for flux calibration, phase calibration and also to determine the bandpass 
shape. Data analysis was done using AIPS. After flagging out bad data, the flux density
scale and instrumental phase were calibrated. The continuum emission was then 
subtracted from the multi-channel visibility data set using the task UVSUB. 
Any residual continuum was then subtracted in the image plane by  fitting a 
linear baseline to line-free regions using IMLIN. 

For the 21-cm emission spectra we took data from the Leiden-Dwingeloo 
survey \citep{h}. The angular resolution of this survey is $\sim 36'$, i.e.
larger than the separation between our target star and the corresponding radio
source. The spectral resolution of the raw data is 1.030 km s$^{-1}$, however 
we Hanning smoothed these spectra to match the resolution of absorption spectra. 

\section[]{Calculations}
\label{temp}
\subsection{Calculation of spin temperature:}

\begin{table*}
\label{table1}
 \caption{Details of our sample and results}
 \begin{tabular}{lcrccccccc}
 \hline
 Star & Radio Source& d$^a$    & Optical & Radio & log(N$_{\rm H_2}$)$^a$ & E(B-V)$^a$ & ~~${\rm T_{01}}^a$ & T$_{\rm s}$ & $\Delta$ T/T \\
 HD   & NVSS        & (pc) &  (l, b) ($^\circ$) & (l, b) ($^\circ$) & cm$^{-2}$          &  mag   & ~K                & K           &\\
 \hline
 22928 &J034227$+$474944&   82 & 150.28, $-$05.77 &  150.20, $-$05.78 &19.30 & 0.01  &~~92~$\pm$~~32&~98~$\pm$~13& $-$0.07 $\pm$ 0.38\\
       &J034629$+$474637&      &                  &  150.76, $-$05.41 &      &       &              &119~$\pm$~04& $-$0.25 $\pm$ 0.35\\
 22951 &J034300$+$340634&  406 & 158.92, $-$16.70 &  158.93, $-$16.51 &20.46 & 0.24  &~~63~$\pm$~~14&~77~$\pm$~02& $-$0.20 $\pm$ 0.22\\
 23408 &J034440$+$243622&   78 & 166.17, $-$23.51 &  165.78, $-$23.51 &19.75 & 0.00  &~~89~$\pm$~~40&103~$\pm$~09& $-$0.14 $\pm$ 0.45\\
 36486 &J053413$-$004408&  384 & 203.86, $-$17.74 &  204.54, $-$17.46 &14.68 & 0.07  &~~1625$\pm$863$^b$ &219~$\pm$~41& $+$1.52 $\pm$ 0.23\\
 36861 &J053450$+$100430&  532 & 195.05, $-$12.00 &  194.89, $-$11.98 &19.11 & 0.12  &~~45~$\pm$~~08&~92~$\pm$~03& $-$0.68 $\pm$ 0.15\\
 37128 &J053550$-$012448&  409 & 205.21, $-$17.24 &  205.36, $-$17.42 &16.57 & 0.08  &~~108~$\pm$~~10$^b$&108~$\pm$~05& $-$0.00 $\pm$ 0.10\\
 38771 &J054409$-$091739&  520 & 214.51, $-$18.50 &  213.75, $-$19.14 &15.68 & 0.07  &~~156~$\pm$~~21$^b$&~72~$\pm$~01& $+$0.74 $\pm$ 0.12\\
 40111 &J055703$+$261119& 1247 & 183.97, $+$00.84 &  183.66, $+$00.77 &19.74 & 0.15  &~117~$\pm$~~52&~88~$\pm$~03& $+$0.28 $\pm$ 0.44\\
 47839 &J064145$+$094704&  705 & 202.94, $+$02.20 &  203.12, $+$02.32 &15.55 & 0.07  &~~1153$\pm$493$^b$ &~54~$\pm$~02& $+$1.82 $\pm$ 0.07\\
 57060 &J071741$-$241542& 1871 & 237.82, $-$05.37 &  237.46, $-$05.43 &15.78 & 0.18  &~~~82~$\pm$~~06$^b$&~72~$\pm$~07& $+$0.12 $\pm$ 0.13\\
 57061 &J071717$-$250453&  933 & 238.18, $-$05.54 &  238.14, $-$05.89 &15.47 & 0.15  &~~513~$\pm$~~46$^b$&147~$\pm$~03& $+$1.11 $\pm$ 0.06\\
 143275&J160052$-$221214&  155 & 350.10, $+$22.49 &  350.51, $+$22.70 &19.41 & 0.16  &~~56~$\pm$~~12&~56~$\pm$~01& $+$0.00 $\pm$ 0.22\\
       &J160401$-$222341&      &                  &  350.92, $+$22.04 &      &       &              &~48~$\pm$~00& $+$0.15 $\pm$ 0.21\\
 147165&J162400$-$261231&  142 & 351.31, $+$17.00 &  351.28, $+$16.11 &19.79 & 0.38  &~~64~$\pm$~~12&~53~$\pm$~03& $+$0.18 $\pm$ 0.19\\
 149757&J163631$-$105841&  138 & 006.28, $+$23.59 &  005.82, $+$23.47 &20.65 & 0.32  &~~54~$\pm$~~04&~80~$\pm$~07& $-$0.39 $\pm$ 0.12\\
 209975&J220320$+$624033& 1086 & 104.87, $+$05.39 &  104.94, $+$05.83 &20.08 & 0.38  &~~77~$\pm$~~21&~67~$\pm$~00& $+$0.14 $\pm$ 0.26\\
 224572&J000020$+$553908&  377 & 115.55, $-$06.36 &  115.72, $-$06.50 &20.23 & 0.17  &~~82~$\pm$~~23&191~$\pm$~05& $-$0.80 $\pm$ 0.24\\
\hline
\end{tabular}\\
\begin{flushleft}
$^a$ \citet{a}; $^b$ Error in T$_{01}$ from our calculations
\end{flushleft}
\end{table*}

    For a homogeneous cloud, the emission and absorption spectra uniquely yield
the spin temperature, i.e.
\begin{equation}
\rmn {T_s=\frac{N(HI)}{1.823\times10^{18}\int\tau(v)\dV}}\,
\label{eqn:tspin}
\end{equation}
with N(H~{\sc i}) being determined from the (off source) emission spectrum and 
$\tau ({\rm v})$ being determined from the absorption spectrum. In the real life 
situation where the gas is not homogeneous, but instead has density and
temperature structure both along as well as transverse to the line of sight, the 
determination of ``the'' spin temperature from radio observations is non trivial. 
As an example, application of Eq.(\ref{eqn:tspin}) to the observed spectrum 
produced by a set of optically thin multiple components along the line of sight
will yield a column density weighted harmonic mean temperature of the individual 
components. If the optical depths are large or there is structure both along and 
transverse to the line of sight, then in general there is no unique interpretation 
of the data, although, several approaches to modeling have been 
attempted \citep[e.g.][]{m,t,i,p}. We note that though there is no mathematically
unique and physically robust procedure to interpret the spectra in this case,
much of what we have learned about the neutral atomic medium has comes from
the relatively simplistic assumptions underlying Eq.~(\ref{eqn:tspin}).
In the analysis presented below, we continue in this tradition.

\begin{figure}
\begin{center}
\includegraphics[angle=-90,width=8.1cm]{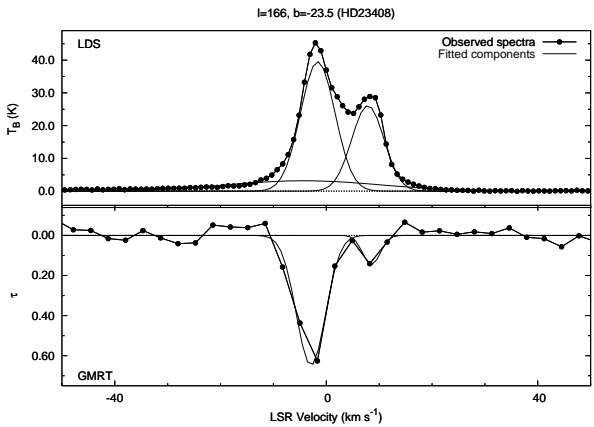}
\end{center}
\caption{Profiles of the observed \HI 21-cm emission and absorption towards l=$166^\circ$, 
b=$-23^\circ.5$ (corresponding to J034440+243622). The top panel is the emission spectrum 
taken from Leiden-Dwingeloo Survey (LDS). The bottom panel is the absorption spectrum 
taken using GMRT. Note that the broad component in emission is not detected in absorption.}
\label{figure1}
\end{figure}

There is in general absorption at several discrete velocity ranges arising from
gas between us and the target star.  As the absorption lines of \htwo in J=0 and J=1 
levels are saturated, it is not possible to obtain the velocity information required to measure \tzeroone 
for individual components; instead what is  measured is the line of sight average value of 
T$_{01}$. It would hence be appropriate to also deal with the average T$_{\rm s}$. 
Accordingly, we first independently decomposed the emission and absorption spectra 
into multiple Gaussian components for the complete available velocity range using 
least-square minimization with minimum number of components. There is of course
some measure of subjectivity in this decomposition. As a check on this, we note that
although the decomposition was done independently for the emission and absorption
spectra, for most cases we found components whose velocities match within the velocity 
resolution in the emission and absorption spectra. In some cases there is one or 
more wide components in the emission spectra which have no corresponding component in 
the absorption spectra (see Fig.~\ref{figure1}). The traditional interpretation of 
this is that they arise from gas with too high a spin temperature to produce measurable 
absorption \citep{n,o,i,p}, i.e. the warm neutral medium (WNM), although of course it
could also arise from a multiple weak narrow components or problems with the spectral
baseline. Given the lack of velocity information in the \htwo spectra, there is no
unique way to match the \HI absorption components found in this way with the \htwo
absorbing gas. We have instead used the velocities of Na~{\sc i}, K~{\sc i} or C~{\sc i} absorption 
lines (taken from \citep*{jen,price,whk,wnh,whi}) towards these same stars as a guide in
this matching process. As the ionization  potential of these species are less than that 
of \HI and close to the energy required to destroy H$_2$, they are expected to coexist 
with \htwo in diffuse ISM. One should note however that (i)~the combination of galactic rotation
and velocity dispersion could cause distant \HI to appear at the same velocity as nearby
gas, (ii)~there could be regions where these ions do not coexist (i.e. the species are
in a higher ionization state) but hydrogen is still in neutral state. Both of these
mean that one cannot make a robust match between the \HI and \htwo absorbing gas.
We do find however that in most of the cases we can find matching components, and
that these components by and large have $ |V_{\rm LSR}| \la 10$~\kms (Given that
the stars are relatively nearby  $ |V_{\rm LSR}| \la 10$~\kms is a plausible
cutoff velocity for gas lying between us and the stars). We therefore proceed
by assuming that this matching is statistically correct. The average \tspin was 
then calculated  from the integrated emission and absorption spectra for all 
the matching CNM components. The calculation is done assuming that one is working in the
limit of small optical depth, and also with no correction for absorption 
of the emission from one cloud by another cloud that happens to lie in
front of it. This is for two reasons (i)~as discussed above, there is no
unique way to make this correction and (ii)~as discussed in more detail
below, we expect all such corrections to lie within the error bars of the 
\tzeroone measurement.  The errors in \tspin were estimated from the rms noise of 
the spectra  and the FWHM of the Gaussian component for emission and 
absorption. This error should be taken only as an indicative value as 
it does not account for the error in fit and the error introduced through 
the assumption of small $\rm{\tau}$. However, since the error bars are
in general small compared to the error bars in T$_{\rm 01}$, the indicative
value should suffice.

\subsection{Calculation of ortho-para temperature:}

It is a standard procedure, in ISM studies, to use measured ortho-to-para
ratio (OPR) of H$_2$ to infer the kinetic temperature of the gas. 
If we assume most of the ortho and para H$_2$ reside in their respective ground 
levels then we can write,
\begin{equation}
\rm {OPR \simeq {N(J=1)\over N(J=0)} = 9\,\exp(-170.5/T_{01})}\ .
\label{eqnopr}
\end{equation}
\tzeroone derived from the above equation will either trace the kinetic 
temperature of the gas (if OPR is controlled by proton or hydrogen exchange 
collisions) or formation temperature (if OPR is controlled by reactions in 
the grain surface) of H$_2$ \citep[for details refer to][]{stern,tak}. We 
have used the observed column densities given in Table~1 and Eq.(\ref{eqnopr}) 
to estimate T$_{\rm 01}$. The associated error is computed using the 
standard error propagation. Since we can not resolve individual components, 
the derived \tzeroone is the average temperature along the line of sight.

\section{Results}
\label{sec:res}

\begin{figure}
\begin{center}
\resizebox*{8.1cm}{8cm}{
\includegraphics[angle=-90,width=8.1cm]{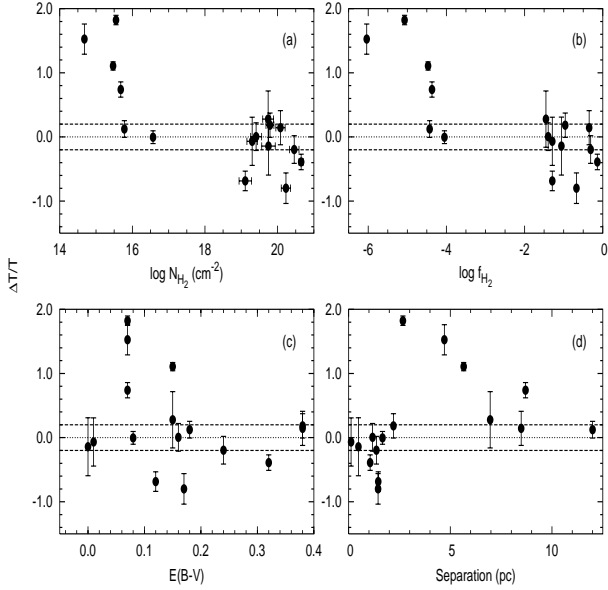}}
\end{center}
\caption{$\rmn{\Delta T/T}$ as a function of
different physical quantities. The horizontal lines in
the figures mark {$\rmn{\Delta T/T} =0$} and 20\% uncertainty.}
\label{figure2}
\end{figure}

We define, the fractional difference between \tspin and \tzeroone by
\begin{equation}
\rm{
\Delta T/T={\frac{2(T_{01}-T_s)}{(T_{01}+T_s)}}
}
\end{equation}
From Table~1 we can see that ${\rm\Delta T/T} = 0$ within measurement 
uncertainties for nine out of 16 cases. For three sightlines we find  \tspin $>$ 
\tzeroone and the other four sightlines have \tspin $<$ T$_{\rm 01}$.
From the Fig.~\ref{figure2} it is also clear that the points
with ${\rm \Delta T/T \simeq 0}$ spread over the whole range of 
$E(B-V)$ and line of sight separation covered in our sample.
However, the points with  ${\rm \Delta T/T > 0}$ mainly come
from the sightlines that have log N(${\rm H_2}$)$\le15.8$ or
molecular fraction f$_{\rm H_2}$(=2N(H$_2$)/2N(H$_2$)+N(H~{\sc i})) 
$\le 10^{-4}$. This is expected as Solomon pumping dominates the
excitation in this regime. The three points with ${\rm \Delta T/T < 0}$ 
are from lines of sight that are optically thick in \htwo with 
f$_{\rm H_2}\ge10^{-2}$ and may be region where \HI and \htwo are not co-existing. 
\begin{table}
\caption{Spatial variation of temperature:}
 \begin{tabular}{@{~~~}lr@{.}lr@{.}lr@{.}lr@{.}lr}
 \hline
{Source} & \multicolumn{4}{c}{Position} & \multicolumn{2}{c}{ds} & \multicolumn{2}{c}{logN(H\sc i)} & T\\
HD & \multicolumn{2}{c}{l ($^\circ$)} & \multicolumn{2}{c}{b ($^\circ$)} & \multicolumn{2}{c}{[pc]} & \multicolumn{2}{c}{(cm$^{-2}$)} & [K]\\
\hline
HD22928  & 150 & 28 & $-$05 & 77 & \multicolumn{2}{c}{---}&\multicolumn{2}{c}{---}&  92\\
radio1   & 150 & 20 & $-$05 & 78 & 0 & 12 &   ~~20 & 87 &  98\\
radio2   & 150 & 76 & $-$05 & 41 & 0 & 86 &   21 & 14 & 119\\
\hline
HD23408  & 166 & 17 & $-$23 & 51 & \multicolumn{2}{c}{---}&\multicolumn{2}{c}{---}&  89\\
radio1   & 165 & 78 & $-$23 & 51 & 0 & 48 &   21 & 00 & 103\\
HD23480  & 166 & 57 & $-$23 & 75 & 0 & 60 &\multicolumn{2}{c}{---}&  67\\
\hline
HD36861  & 195 & 05 & $-$12 & 00 & \multicolumn{2}{c}{---}&   20 & 78 &  45\\
radio1   & 194 & 89 & $-$11 & 98 & 1 & 45 &   20 & 68 &  92\\
HD36822  & 195 & 40 & $-$12 & 29 & 4 & 16 &   20 & 81 &  63\\
\hline
HD37128  & 205 & 21 & $-$17 & 24 & \multicolumn{2}{c}{---}&   20 & 45 & 108\\
radio1   & 205 & 36 & $-$17 & 42 & 1 & 67 &   20 & 92 & 108\\
HD37742  & 206 & 45 & $-$16 & 59 & 9 & 65 &   20 & 41 & 101\\
\hline
HD57060  & 237 & 82 & $-$05 & 37 & \multicolumn{2}{c}{---}&   20 & 70 &  82\\
radio1   & 237 & 46 & $-$05 & 43 &12 & 00 &   20 & 51 &  72\\
HD57061  & 238 & 18 & $-$05 & 54 &12 & 95 &   20 & 70 & 513\\
radio2   & 238 & 14 & $-$05 & 89 &19 & 89 &   20 & 24 & 147\\
\hline
HD143275 & 350 & 10 & $+$22 & 49 & \multicolumn{2}{c}{---}&   21 & 15 &  56\\
radio1   & 350 & 51 & $+$22 & 70 & 1 & 17 &   21 & 09 &  56\\
radio2   & 350 & 92 & $+$22 & 04 & 2 & 38 &   20 & 60 &  48\\
\hline
\end{tabular}
\label{table2}
\end{table}
In panel (d) we plot $\Delta$ T/T as a function of maximum
separation between optical and radio sightlines. We do not find
any clear trend in this plot.
We explore this issue further by collating in Table~2 
the cases where we have more than 1 radio source or star within 
a separation of $1^\circ$. The columns in the table are
(1)~the name of the sources, (2)~position, (3)~separation
of the lines of sight (ds) at the distance of the star, 
(4)~N(H~{\sc i}) and (5)~the temperature (\tzeroone or T$_{\rm s}$) 
measured along the line of sight. We find, among the cases
where \tzeroone $\simeq$ T$_{\rm s}$, the match generally
improves with decreasing separation, though there are
cases (e.g. HD36861) where the temperatures do not match 
even for a separation of 1.45~pc. One should note however
that the separations are computed assuming the gas is 
at the distance of the star, while in reality the gas could
lie anywhere between us and the star.

\section{Discussion}
\label{sec:dis}

In this work we present the GMRT measurement of 21-cm spin 
temperature toward 18 radio sources that are close to 16 bright starts 
for which UV observations are available. We find the \tspin and 
\tzeroone trace each other within the observational uncertainties 
when N(H$_2$)$\ge10^{15.8}$ cm$^{-2}$ in 75 per cent of the cases. 
\tzeroone is found to be higher than \tspin for the sightlines
with low \htwo column density. 

\citet{i} have performed similar analysis like us
towards three directions. There is one star in our sample 
(HD~22951) that is in  common with \citet{i}. For this sightline 
\tzeroone = 63$\pm$14 K and we find consistent 
\tspin = 77$\pm$2 toward a radio  source with a maximum separation 
of 1.35 pc. \citet{i} report \tzeroone = 27$\pm$13 and 29$\pm$11 K 
toward NRAO 140 and 3C93.1 that are separated by 14.6 and 9.5 pc 
respectively from the star. The spin
temperature that they quote is that for a single CNM component 
along the line of
sight. If we instead use the column density weighted average of 
all CNM components in the velocity range probed by the C~{\sc i} absorption \citep{jen}, 
the spin temperatures are 53 and 72~K, i.e. in agreement with T$_{\rm 01}$. 

A major concern is that we do not know if our temperature comparisons are 
for the same gas; we have only a plausible matching between the \HI and \htwo
absorbing gas. As discussed above though, this matching is likely to be
on the average correct. If this is so, then our findings mean that \tspin and \tzeroone 
do not vary much over distances of a few parsecs. This is in apparent contrast to the well
known findings that the \HI 21-cm opacity varies on much smaller spatial scales \citep*[e.g.][]{l,x}.
The resolution may be that the opacity fluctuations reflect not so much fine
scale structure in the temperature as fine scale structure in the velocity
and density fields \citep*[e.g.][]{b}. It is also possible that our averaging
over the several line of sight components decreases the effects of small
scale variations.

Another issue is related to the relatively large error bars in T$_{\rm 01}$. Is it possible
that all that our data is telling us is that the CNM has a characteristic
temperature $\sim 80$~K, and that the general agreement between \tzeroone
and \tspin merely reflects the fact that both the UV and radio observations
are probing the CNM? Or does
the temperature agreement actually extend to individual lines of sight, 
as we have been asserting? A least-square linear fit using all 11 lines of sight for which 
\tzeroone and \tspin agree (including 
both the radio sources close to HD22928 and HD143275) for \tzeroone$={\rm r}\,$\tspin gives r$=0.981
\pm0.054$. For all the lines of sight with H$_2$ column density higher than our threshold, 
the least-square fit gives r$=0.778\pm0.080$. As a quantitative check, we have calculated 
the Spearman correlation coefficient for increasingly larger subsamples. For a subsample with 
all 11 lines of sight for which \tspin matches T$_{\rm 01}$, the correlation coefficient is 
maximum (0.648). The corresponding significance is $p < 0.05$ (from a two-tailed test). 
For a subsample with all 14 lines of sight for which \htwo column density is more than 
10$^{15.8}$ cm$^{-2}$, the correlation coefficient is 0.545. On the other hand when one includes
lines of sight with N(H$_2$) less than 10$^{16}$ cm$^{-2}$, the correlation coefficient 
goes down to 0.406 (for our complete sample). To summarize then, as per the Spearman rank 
coefficient test, at the better than 95\% level, there is a one
to one relation between \tzeroone and \tspin for the 11 lines of sight with N(H$_2$) 
higher than $10^{15.8}$ cm$^{-2}$. Agreement between \tspin and \tzeroone for high 
column density gas would mean that the ortho-para equilibrium is mainly due to 
exchange collisions and that T$_{01}$ does not reflect the formation temperature. 
The absence of relation between the two temperatures when N(H$_2$) is optically thin
is consistent with the slow rate of exchange collisions compared to the
H$_2$ destruction rate and excess Solomon pumping from J=0 and J=1 levels that make 
\tzeroone deviate from the kinetic temperature and hence T$_{\rm s}$.

\section*{Acknowledgments}

The observations presented in this paper were obtained using the GMRT which
is operated by the National Centre for Radio Astrophysics (NCRA) of the Tata 
Institute of Fundamental Research (TIFR), India. We are grateful to the anonymous
referee for prompting us into substantially improving this paper. Some of the data presented 
in this paper were obtained from the Multimission Archive at the Space 
Telescope Science Institute (MAST).

\bsp

\label{lastpage}

\end{document}